\documentclass{article}

\usepackage{arxiv}
\usepackage[utf8]{inputenc} % allow utf-8 input
\usepackage[T1]{fontenc}    % use 8-bit T1 fonts
\usepackage{hyperref}       % hyperlinks
\usepackage{url}            % simple URL typesetting
\usepackage{booktabs}       % professional-quality tables
\usepackage{amsfonts}       % blackboard math symbols
\usepackage{nicefrac}       % compact symbols for 1/2, etc.
\usepackage{microtype}      % microtypography
\usepackage{lipsum}
\usepackage{fancyhdr}       % header
\usepackage{graphicx}       % graphics
\usepackage{amsmath}
\graphicspath{{Plots/}}     % organize your images and other figures under Plots/ folder

\newcommand{\indep}{\perp\! \! \! \perp}

\title{Interpretational challenges of the Win Ratio in analyzing Hierarchical Composite Endpoints in Chronic Kidney Disease}

  \author{Henrik F. Thomsen \\
    Biometrics CKAD 2, Biometrics CKAD, Novo Nordisk A/S, \\ 
    Aalborg, Denmark \\
     \texttt{hfth@novonordisk.com} \\ 
     \and 
    Samvel B. Gasparyan \\
    Late-Stage Development, Cardiovascular, Renal and Metabolism (CVRM), \\ 
    BioPharmaceuticals R\&D AstraZeneca, Boston, Massachusetts, USA \\
    \and
    Julie F. Furberg \\
    Biometrics CKAD 1, Biometrics CKAD, Novo Nordisk A/S, \\ 
    Søborg, Denmark \\   
    \and
    Christoph Tasto \\
    Evidence Generation and Decision Science, Sanofi, \\
    Frankfurt am Main, Germany \\
    \and 
    Nicole Rethemeier \\
    Clinical Statistics \& Analytics, Bayer AG, \\
    Wuppertal, Germany \\
    \and 
    Patrick Schloemer \\
    Clinical Statistics \& Analytics, Bayer AG, \\
    Berlin, Germany \\
    \and
    Tuo Wang \\
    Global Statistical Science, Eli Lilly and Company, \\
    Indiana, USA \\
    \and 
    Niels Jongs \\
    Department of Clinical Pharmacy and Pharmacology, \\ 
    University Medical Center Groningen, Netherlands \\
    \and 
    Yu Du \\
    Global Statistical Science, Eli Lilly and Company, \\
    Indiana, USA \\ 
    \and \\
    Tom Greene \\
    Department of Internal Medicine, University of Utah, \\
   Salt Lake City, Utah, USA     
    }

\begin{document}

\maketitle

\begin{abstract}
Win statistics based methods have gained traction as a method for analyzing Hierarchical Composite Endpoints (HCEs) in randomized clinical trials, particularly in cardiovascular and kidney disease research. HCEs offer several key advantages, including increased statistical power, mitigation of competing risks, and hierarchical ranking of clinical outcomes.  
While, as summary measures, the win ratio (WR) along with the Net Benefit (NB) and the Win Odds (WO) provide a structured approach to analyzing HCEs, several concerns regarding their interpretability remain. In this paper, we present known issues with the WR using simple examples designed to explore the implications for the clinical interpretability of the treatment effect measure in the chronic kidney disease setting. Specifically, we discuss the challenge of defining an appropriate estimand in the context of HCEs using the WR, the difficulties in formulating a relevant causal question underlying the WR, and the dependency of the WR on the variance of its components, which complicates its role as an effect measure. Additionally, we highlight the non-collapsibility and non-transitivity of the WR, further complicating its interpretation. While the WR remains a valuable tool in clinical trials, its inherent limitations must be acknowledged to ensure its proper use in regulatory and clinical decision-making.
\end{abstract}


\section{Introduction}\label{intro}

In recent years, there has been a notable surge in the analysis of Hierarchical Composite Endpoints (HCE) using the Win Ratio (WR) in randomized clinical trials. Mao's study \cite{mao2024defining}, based on a search of ClinicalTrials.gov, highlights a significant rise in the application of WR analyses. 
Within the realm of cardiovascular and kidney disease, several major outcome trials have employed the WR in some capacity to compare a novel treatment intervention with a comparator \cite{Pocock2023-mp, Gillmore2024-fp, Heerspink2023-gt, Little2023-cq, kosiborod2024semaglutide, gasparyan2022design, maurer2018tafamidis}.

The application of the Win Ratio (WR) in clinical trials has drawn critiques from a clinical perspective; see, for example, \cite{Ajufo2023-zb} and the discussion by \cite{Davison2025-fc, Packer2025-ve}. Our aim in this article is to present already established statistical issues with the WR from a perspective of clinical interpretability, using simple examples designed to enhance understanding. For a comprehensive overview of the general pairwise comparison approach to data analysis, consult \cite{Buyse2025-ow}. As context for our discussion, we focus specifically on two-armed clinical trials evaluating the efficacy of an active treatment against a comparator.

\subsection{Hierarchical composite endpoint}
A common feature of the HCEs used in almost all of the above-mentioned trials is that they typically involve one terminal event, one or more non-terminal events, and one or more continuous/ordinal outcomes. For example, \cite{Pocock2023-mp}, defined a heart failure HCE as ".\textit{..a hierarchical composite of death, number of HF events, time to first HF event, or a $\geq$ 5-point difference in Kansas City Cardiomyopathy Questionnaire (KCCQ) total symptom score change at 90 days}." Similarly, \cite{Heerspink2023-gt} proposed and investigated a kidney disease HCE comprising "\textit{(1) all-cause mortality; (2) kidney replacement therapy defined as dialysis for at least 28 days or kidney transplantation; (3) sustained GFR <15 mL/min per 1.73 $\text{m}^2$ for at least 28 days; (4) sustained GFR decline from baseline for at least 28 days of 57\%; (5) 50\%; (6) 40\%; or (7) GFR slope}". GFR slope refers to the individual-level, annualized rate of decline in glomerular filtration rate (GFR), which represents kidney function. 

HCEs allow for a straightforward combination of evidence from different types of measurements (e.g., time-to-event, ordinal, and continuous). Apart from this, there are several further compelling reasons for utilizing HCEs. First, HCEs acknowledge the varying importance of different components by establishing a hierarchy among the components. Second, incorporating terminal events helps in mitigating competing risks. Third, employing HCEs can in some cases lead to increased statistical power compared to analyzing individual components separately or solely focusing on the time to the first occurrence of any of the event-components. Furthermore, as with other types of composite endpoints, using HCEs eliminates the necessity for making adjustments for multiplicity to control type I error, which is typically required when analyzing the individual components separately.

While the ordering of the HCE may seem straightforward, it is not always the case. To illustrate this with an extreme example, suppose a trial where all subjects in the comparator arm systematically experience the terminal event slightly earlier than those in the active arm. At the same time, all active arm  subjects experience a non-terminal severe event within the HCE, and none in the comparator arm do. In this case, win statistics based on the HCE will indicate superiority for the active arm due to the small benefit on the terminal event, yet the HCE ignores the increased frequency of non-terminal severe events in the active arm. These limitations are also present in other analyses of composite endpoints; for example, a time-to-first-event analysis of this type of data would encounter similar issues. To address these concerns, it is in general advisable to also evaluate the components individually \cite{FDAmult, EMAmult}
 
\subsection{Win Ratio}
The WR, as proposed by \cite{Pocock2012-mo}, serves as a summary measure for analyzing HCEs when comparing two treatment arms. In the unmatched version of this approach, each subject from the novel treatment arm is compared to each subject from the comparator arm to determine a "winner" for each comparison. Note that although "success" might be considered a more appropriate term in some contexts, the use of "winner" is well established in the literature. Initially, the analysis determines whether one subject wins over the other on the first HCE component. If a clear winner emerges, that subject is declared the winner for that specific subject comparison. If a winner cannot be determined, a tie is declared for this component, and the process moves on to the next component. This sequence continues until a winner is declared for the entire comparison or until a tie is declared for the last component, resulting in an overall tie for that particular subject comparison. Let $P_t$ denote the proportion of pairs when treatment wins, $P_c$ denote the proportion of pairs when control wins, and $P_\text{tie} = 1-P_t-P_c$ denote the proportion of ties. The WR is then estimated as $P_t / P_c$.  Consequently, the win ratio does not account for ties, and in situations where the proportion of ties is large, it may provide a misleading measure, similar to the risk ratio in analyses of dichotomous endpoints when absolute risks are small.
In addition to concerns about ties, another important consideration in the application of the WR is how to account for baseline covariates that may influence outcomes. Covariate adjustment of the win ratio is an area in development, see e.g. \cite{WRGasparyan}; a simple, often-employed method is stratification. The stratified win ratio considers predefined strata and is calculated by comparing the number of wins for each treatment group within each stratum, then aggregating these results with appropriate weights.

Other often used summary measures for HCEs are the net benefit (NB) and the Win Odds (WO). The NB, also referred to as the Win Difference, computes the difference in win probabilities between treatment arms, i.e. $P_t - P_c$. The WO, estimated as $(P_t + 0.5P_\text{tie})/(P_c + 0.5P_\text{tie})$, adjusts for tied outcomes by allocating half of the ties to the wins for each arm, and thus 'penalizing' trials with many ties. While these three win statistics differ in their interpretations and sensitivity to ties, their corresponding statistical tests are asymptotically equivalent under standard regularity conditions \cite{dong2023win, dong2024approximate}. We will focus on the WR in this article, but briefly mention NB and WO when relevant. 

All of the benefits and issues presented in this section are to be considered when applying a WR to HCEs. However, in this article, we will direct our focus toward specific aspects related to the interpretation of the WR estimate. First, we will delve into the challenge of defining a suitable estimand for which a WR analysis of an HCE would provide a proper estimator, particularly formulating the clinical question that the WR would aptly address. Second, we will tackle the difficulty of defining a relevant causal question that can be answered by a WR analysis. Third, we also discuss the non-transitivity of the WR. Fourth, we will address the fact that a WR of an HCE with one or more continuous components will be contingent on the variance of those components.  As a consequence, the WR is also a non-collapsible measure (similar to hazard or odds ratios).  Taken together, these considerations add significant complexity to the interpretation of the WR as an effect measure 

All these factors collectively lead to the acknowledgment that the win ratio is problematic as an effect measure, warranting further methodological refinement.
 
\section{Estimand}\label{estimand}

The estimand framework was introduced in 2020 with the publication of the ICH E9 addendum \cite{ICHE9Addendum}. This framework provides a structured approach to defining estimands in clinical trials, ensuring clarity, transparency, and interpretability of clinical trial results, thereby improving decision-making processes in drug development and regulatory assessments. The key attributes of the estimand include: 
\begin{itemize}
    \item[$ $]\textbf{Treatment}: Defines the treatment and alternative treatment of interest, including choices of standard-of-care.
    \item[$ $] \textbf{Population}: Identifies the group of subjects relevant to the clinical question.
    \item[$ $] \textbf{Variable (Endpoint)}: Specifies the measurement or outcome used to address the clinical question.
    \item[$ $] \textbf{Intercurrent Event Strategy}: Describes how intercurrent events are accounted for,  e.g. through treatment policy, hypothetical, or composite strategies.
    \item[$ $] \textbf{Population-level Summary}: Details the summary measure of the variable across the population.
\end{itemize}
\noindent The addendum highlights that a key focus in drug development is demonstrating the existence of treatment effects and quantifying their magnitude. This involves the causal comparison of the outcome with the intervention to the outcome that would have occurred for the same subjects under an alternative intervention (i.e., had they not received the treatment or had they received a different treatment). An estimand is a precise description of the treatment effect that reflects the clinical question posed by a given clinical trial objective. 

Therefore, it is crucial to clearly specify the clinical question being addressed by the trial and define a measure of the treatment effect that quantifies this. 

It is important to acknowledge that, from a strict estimand perspective,  considerations of power or technical difficulties related to specific statistical analyses should not influence the definition of the estimand. For a WR analysis to be chosen, one must argue that assessing wins is an appropriate way to answer the clinical question.

For example, in a chronic kidney disease (CKD) trial using an HCE consisting of all-cause mortality; kidney replacement therapy; and GFR slope, it might be reasonable to argue that a better outcome defined by this HCE (a win) is a suitable endpoint. Thus, one could define the estimand as:  
\begin{itemize}
  \item[$ $]\textbf{Treatment}: Assigned novel or alternative treatment, on top of standard of care.
   \item[$ $]\textbf{Population}: CKD population of interest. 
   \item[$ $]\textbf{Variable (Endpoint)}: Time to the most severe of the first \textit{n-1} components within \textit{t} months. If none of the time-to-event components occur within \textit{t} months, difference in GFR slope at \textit{t} months is considered.
   \item[$ $]\textbf{Intercurrent Event Strategy}: Relevant intercurrent events like death, dialysis, or transplantation, are part of the endpoint and are handled with a composite strategy.
   \item[$ $]\textbf{Population-level Summary}: The ratio of the probability that a random subject assigned to the intervention has a better outcome to the probability that a random subject assigned to intervention has a worse outcome.
\end{itemize}
\noindent Based on this one could consider the HCE analyzed with WR as a holistic assessment of kidney function change, giving more emphasis on more severe kidney events. In section~\ref{causal} we will discuss the causal interpretation of the WR as the population-level summary. 

A fundamental consideration when defining an estimand for a WR-based analysis is its dependency on the maximum follow-up time and the censoring distribution. Since the WR is based on pairwise comparisons between subjects, its estimation is affected by the time window over which these comparisons are conducted~ \cite{oakes2016win}. To define a valid estimand that is independent of censoring, one approach is to restrict WR calculations to a fixed time horizon $t$, such as 24 months \cite{wang2024restricted, dong2020inverse}, ensuring that any observed differences in follow-up arise exclusively from loss-to-follow-up. This reduces interpretational complexities from differential follow-up, as  the contributions of the more severe endpoints comprising the HCE tend to increase with longer follow-up time; thus, the relative contributions of the different endpoints will generally vary between patients with shorter versus longer follow-up durations. It also improves comparability across studies, provided that the fixed follow-up period is identical in each study 

Alternatively, if clinically justified, the WR can be assumed to remain constant over time, allowing estimation of a single value representing the entire study period \cite{mao2021class, wang2022stratified}. Despite these considerations, testing H$_0$ : WR = 1 remains valid, even if its interpretation depends on follow-up duration. The usual statistical tests for WR, such as asymptotic methods or permutation-based approaches, remain robust for evaluating the null hypothesis of no treatment effect. However, careful interpretation of the WR estimate is necessary to ensure it aligns with the clinical question of interest.

If a treatment on average leads to a higher win probability compared to a comparator, this indicates the existence of a positive treatment effect. However, it is more complex to claim that the estimated WR also answers the "how much better" question. As mentioned in the introduction, wins can be qualitatively different (e.g., longer survival or a difference in GFR slope). For this reason, as well as for a greater depth of understanding of the treatment’s effects, analyses of the components of the composite endpoint are important and can influence interpretation of the overall results \cite{FDAmult}.  This situation mirrors other composite endpoints, but it can be particularly pronounced for HCEs because their hierarchical structure broadens the scope of outcomes that can be included in the composite. For instance, a large WR driven by differences in GFR slope does not carry the same implications as a WR of the same magnitude driven by differences in survival. 

If the primary clinical interest lies in comparing the location of a continuous endpoint between treatment arms, in a population where a significant number of deaths is anticipated during the trial, one might consider employing an HCE with time to death as the first component, and the continuous endpoint as the second in order to circumvent interpretational challenges resulting from early termination of follow-up by "truncation by death", thus, in this context, the clinical event of death act as nuisances that complicate the evaluation of the continuous endpoint. However, if we are to remain faithful to the strict estimand framework, the technical challenges of managing intercurrent events, such as death, in the comparison of continuous outcomes  should not dictate the choice of the estimand. Specifically, if the estimand aims to assess the treatment effect on the continuous endpoint, the composite strategy of handling intercurrent events through an HCE will not adequately address the estimand of interest. Moreover, if applied to the normally or log-normally distributed continuous endpoint, the WR as a treatment effect measure does not estimate the difference in the locations of the continuous distributions (as can be seen from the example below of normally distributed random variables, WR is in this case related to the \textit{standardized} mean difference, see equation~\eqref{eq:winprobnorm} and not the mean difference); rather, it assesses the odds that a randomly selected subject assigned to the intervention has a better outcome than a randomly selected subject assigned to the control group. 

On the other hand, in the CKD example the primary clinical interest lies in the treatment effect on kidney function decline. Thus endpoint components of interest would be GFR slope and relevant clinical events such as dialysis or kidney transplantation. In this setting, these clinical events are essential and intentionally incorporated into the composite endpoint, as they provide crucial insights into kidney function. Interpreting GFR values after dialysis or transplantation is not straightforward. Hence, the estimation of GFR slope in the presence of these intercurrent events, which are by themselves highly relevant for the evaluation of kidney function, may also be challenging. To overcome these issues, employing the HCE for the estimand as outlined in the CKD example above might be an attempt for a holistic assessment of kidney function.

\subsection{Causal interpretation}\label{causal}

As mentioned in section \ref{estimand}, the ICH E9 addendum on estimands emphasizes the importance of clearly defining the treatment effect of interest in clinical trials. The addendum encourages the use of causal language and frameworks to describe these effects, ensuring that the treatment effect is interpretable in a causal manner.  For an introductory text on causal inference please refer to \cite{hernan2020}.

In general, causal effect measures can be divided into individual-level and population-level causal effect measures. The individual-level causal effect measure compares potential outcomes on each of two treatment arms for the \textbf{same} individual and then summarizes those comparisons across a population. The population-level causal effect operates with an effect measure at the population level. The difference in means effect measure is both an individual-level and population-level effect measure, i.e. the mean of differences and difference of means are the same. The WR falls into the category of a population-level effect measure. Using potential outcomes notation, the WR can be described as

\noindent 
$$\text{WR} = \frac{\sum_{i,j}P\left[ Y_i(1) \succ Y_j(0) \right] }{\sum_{i,j} P\left[ Y_i(1) \prec Y_j(0) \right]  },$$
where the sum is build over all subjects $i$ and $j$, the $\succ$ operator indicates a 'better' outcome, and $Y(1)$ and $Y(0)$ are the factual/counterfactual outcomes for treatments $1$ and $0$. So $Y_i(1) \succ Y_j(0)$ if subject $i$ has a better outcome on treatment $1$ than subject $j$ on treatment $0$. This measure is in general not the same as the individual-level effect measure.  A corresponding individual-level effect measure would be defined as 

$$\widetilde{\text{WR}} = \frac{\sum_{i}P\left[ Y_i(1) \succ Y_i(0) \right] }{\sum_{i} P\left[ Y_i(1) \prec Y_i(0) \right]  },
$$

\noindent where comparisons are done within subjects. However this is not identifiable in a randomized experiment \cite{CausWilcoxFay, mao2018causal}. 

There is no guarantee that these two are similar. One can even construct examples with opposing conclusions. As an illustration, consult the so-called Hands paradox \cite{Hand1992-jp, CausWilcoxFay}: in Table~\ref{tab:handsparadox}, we present the factual and counterfactual responses for three subjects under two treatments. Here, Y(0) represents the response under the comparator treatment, while Y(1) denotes the response under the active treatment. A higher response value indicates a better outcome. In this toy example, the WR (population-level), comparing all outcomes in the two arms is: ((0+0+0) + (0+1+0) + (0+1+1)) / ((1+1+1) + (1+0+1) + (1+0+0)) = 3/6 = 0.5, whereas the $\widetilde{\text{WR}}$ (individual-level), doing the comparisons within individuals  (0+1+1)/(1+0+0) = 2/1 = 2. Thus, not only are they different, but they are pointing in opposite directions; one would lead to a conclusion of a beneficial individual-level treatment effect, whereas the population-level would suggest a harmful treatment effect. 

\begin{table}[ht]
\centering
    \begin{tabular}{lcc} 
     \toprule
         \textbf{Subject}&  \textbf{Y(1)} & \textbf{Y(0)} \\ 
         \midrule
         1&  1& 6\\
         2&  3& 2\\ 
         3&  5& 4\\ 
    \bottomrule
    \end{tabular}
    \caption{Toy example with factual and counterfactual outcomes for three subjects}
    \label{tab:handsparadox}
\end{table}

The non-identifiable individual-level $\widetilde{\text{WR}}$ has a straightforward interpretation for patients and physicians: "What is the ratio of the probability of obtaining a better outcome to the probability of a worse outcome when comparing treatment 1 with treatment 0". However, the interpretation of the population-level $\text{WR}$ for a subject/physician is more dubious: "what is the ratio of the probability of obtaining a better outcome to the probability of a worse outcome for a subject exposed to treatment 1 vs another subject exposed to treatment 0". While this population-level comparison is valid in a causal sense, it is not necessarily the causal interpretation relevant to a specific clinical question. This is particularly important when making treatment recommendations at an individual level, where clinicians and patients may be more interested in understanding how a given treatment would affect a specific patient rather than an average comparison across populations. 
The same interpretational concerns apply to the WO and NB, as they are fundamentally based on the same win probability comparisons as the WR. A similar exposition is given in \cite{Verbeeck2025-2}. 
This distinction between individual-level and population-level causal interpretations has been explored in other works as well, including a detailed exposition by \cite{Verbeeck2025-2}. 

\section{Non-transitivity}
A natural expectation from any effect measure is that it adheres to the principle of transitivity. Transitivity implies that if treatment A is deemed superior to treatment B, and treatment B is deemed superior to treatment C, then one should logically conclude that treatment A is superior to treatment C. This property provides coherence and consistency in comparative analyses. Lack of this property would render the effect measure a possibly unreliable tool for decision-making, as it not only complicates clinical decision-making but also limit the effectiveness of meta-analytic approaches, where cross-trial consistency is critical.

Non-transitivity may arise in WR, WO, and NB calculations due to censoring of time-to-event components. This issue occurs when subjects are observed over varying time intervals, necessitating comparisons based on overlapping observation periods. For further details, see \cite{oakes2016win}, and \cite{Oakes2025-sq}. 

Non-transitivity can also occur for continuous or ordinal components. For example, consider three treatments (A, B, and C) with outcomes shown in Table~\ref{tab:nontransitivity}. When comparing treatment pairs using the WR, we observe that WR(A vs B) = WR(B vs C) = WR(C vs A) = $1.25$, which violates transitivity, in  other words, no single treatment emerges as the best because the treatments each outperform each other in a circular manner. This example corresponds to a simplified version of Efron’s dice. A similar argument for the Wilcoxon–Mann–Whitney test is given in \cite{Thangavelu2007-yz}. 

Interestingly, non-transitivity persists even when considering the non-identifiable individual-level effect measure $\widetilde{\text{WR}}$, as illustrated by the example in Table~\ref{tab:nontransitivity}. When the data in the table is interpreted as factual and counterfactual outcomes for three hypothetical subjects, the calculated  $\widetilde{\text{WR}}$(A vs B) = $\widetilde{\text{WR}}$(B vs C) = $\widetilde{\text{WR}}$(C vs A) = $2$. 
Importantly, the challenges discussed here are not exclusive to the WR but extend to the WO and NB as well. 

 Table~\ref{tab:nontransitivity}.

\begin{table}[ht]
\centering
    \begin{tabular}{lccc} 
     \toprule
         \textbf{Subject}&  \textbf{A} & B& \textbf{C} \\ 
         \midrule
        1 & 2 & 1 & 3 \\
        2 & 4 & 6 & 5 \\
        3 & 9 & 8 & 7 \\
    \bottomrule
    \end{tabular}
    \caption{Toy example with outcomes across three treatment arms.}
    \label{tab:nontransitivity}
\end{table}

\section{Variance dependence}\label{var}
In this section, we consider the impact of the variance of a single normally distributed continuous random variable on the win statistics when there is only one continuous component present in the HCE. This simplification helps to clearly illustrate the concept, although the effects will persist even when more components are included, albeit potentially diluted.

\subsection{The impact of the overall variance}
\begin{figure}
    \centering
    \includegraphics[width=1\linewidth]{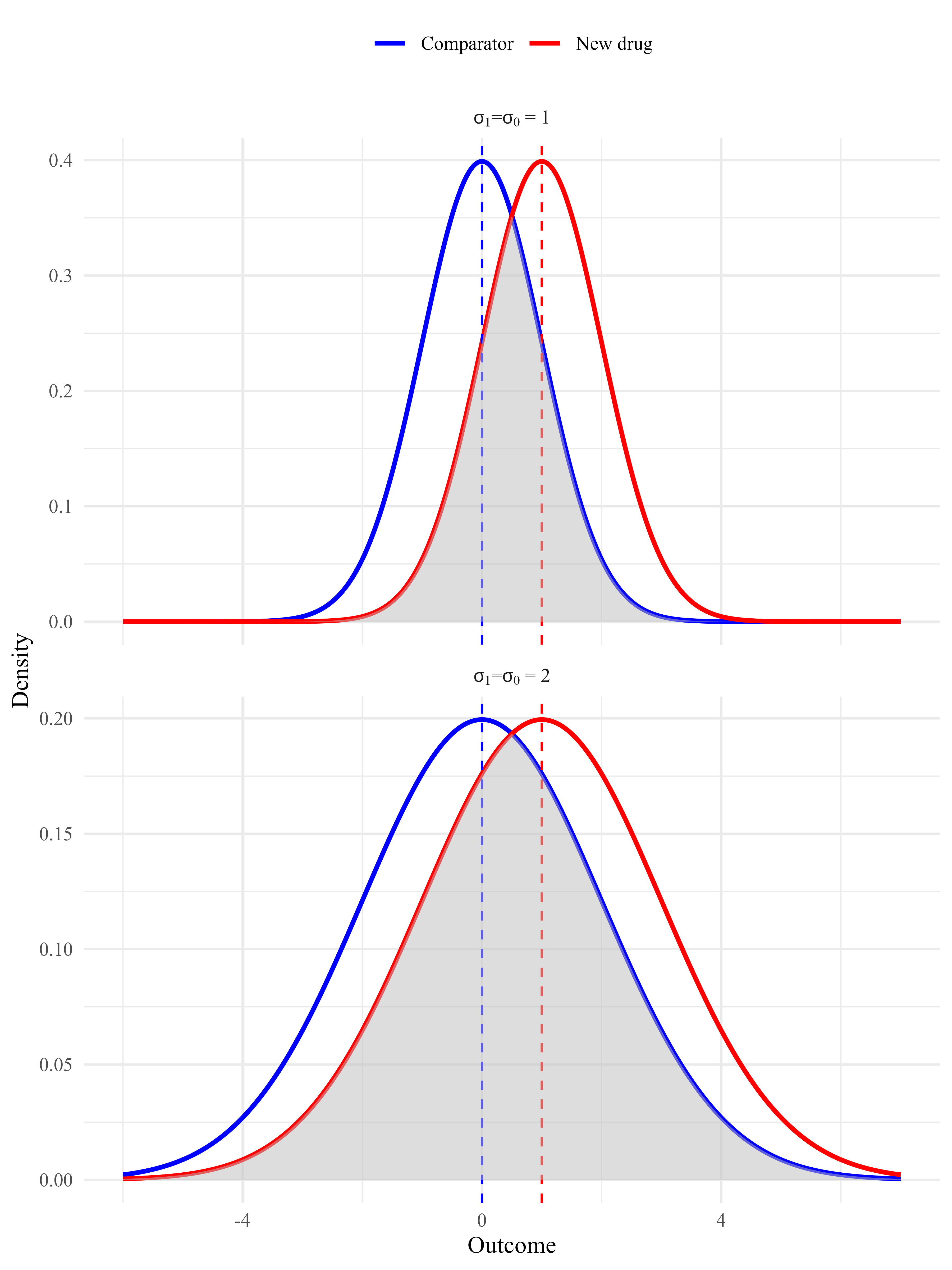}
    \caption{Normal density plots, with a difference of means of 1 and standard deviations of 1 and 2 respectively}
    \label{fig:pdf-diff-var}
\end{figure}

Let $Y_0 \sim \mathcal{N}(\mu_0, \sigma_0^2)$ and $Y_1 \sim \mathcal{N}(\mu_1, \sigma_1^2)$ be two independent normally distributed random variables, where $Y_i$ denotes the response under treatment $i=0,1$. Therefore, given that $Y_1 \indep Y_0$ we have that $(Y_1- Y_0) \sim \mathcal{N}(\mu_1 - \mu_0, \sigma_1^2 + \sigma_0^2).$ If larger values are 'better', then the win probability of $Y_1$  vs. $Y_0$ (see for example \cite{WRGasparyan}) is
\begin{equation} \label{eq:winprobnorm}
    \theta=P(Y_1 >Y_0)=\Phi\left(\frac{\mu_1 - \mu_0}{\sqrt{\sigma_1^2+\sigma_0^2}}\right).
\end{equation}
    Here $\Phi(\cdot)$ is the cumulative standard normal distribution function. One can view
$$\delta=\frac{\mu_1 - \mu_0}{\sqrt{\sigma_1^2+\sigma_0^2}}$$
as the theoretical version of Cohen's d \cite{cohen2013statistical}. Hence, in this case, the win probability is a standardized effect measure as opposed to non-standardized effect measures such as difference in means or medians. Consequently, win statistics are rather a measure of discrimination between two distributions. In the case of continuous distributions, the probability of a tie is 0, and thus the WR and WO are the same and are defined as 
    $$
       \text{WR}= \text{WO} = \frac{\theta}{1-\theta}.
    $$ 
Similarly, the NB is defined as $\nu=2\theta-1$. Therefore, all win statistics depend on the variance of the underlying distributions in case of normally distributed random variables. One can easily show that this is the case for log-normally distributed random variables as well (or any distributions with monotonic transformations of normally distributed random variables). As a consequence of this, the magnitude of the treatment effect estimated by win statistics reflects both the difference in means and the variability of the underlying distributions. Therefore, if  $\sigma=\sqrt{\sigma_0^2 + \sigma_1^2}$, then  $\lim_{\sigma \to\infty} \text{WR}=\lim_{\sigma \to\infty} \text{WO} = 1,$ irrespective of $\mu_0$ and $\mu_1$. If  $\mu_1 > \mu_0$  then we have that $\lim_{\sigma \to 0} \text{WR}=\lim_{\sigma \to 0} \text{WO} = \infty$ (similarly, if $\mu_1 < \mu_0$  then $\lim_{\sigma \to 0} \text{WR}=\lim_{\sigma \to 0} \text{WO} = 0$). One can obtain any WR or WO value from $1$ to $\infty$ solely by altering the variances. For example, in case of a mean difference of 1, for common standard deviations $\sigma_1=\sigma_0$ of 1 and 2, the WO and WR values will be 3.17 and 1.76 respectively. This is illustrated in~Figure~\ref{fig:pdf-diff-var}.  The overlapping of the distributions increases with increasing variance leading to a decrease in the WR. 

The WR's dependency on variation, particularly when measurement error is a substantial contributor to the variation in a continuous variable, closely parallels the scenario encountered when dichotomizing continuous endpoints for 'responder analysis.'. This is thoroughly examined in \cite{responderabugov}, where they conclude that  "\textit{In other words, large responder effect benefits do not necessarily imply large clinical benefits and small responder effect risks do not necessarily imply minimal clinical risk. A large responder benefit may reflect an unimportant clinical benefit amplified by a small standard deviation, and a small responder safety risk may reflect a clinically important change minimized either by a large standard deviation... }". Similarly, the draft FDA guidance on obesity and overweight \cite{FDAguidobese} reflects this concern, stating that "\textit{...responder analyses for the evaluation of weight reduction are generally not recommended."}.

Note that this is a property of the theoretical win statistics, that is, the estimand itself.  In what follows, we will explore the impact of different sources of variability on the WR, including impact of the variability of the measurement error. A similar exposition can be found in \cite{Burzykowski2025-7}. 

\subsection{Slope example}
Let us return to the CKD example from Section~\ref{estimand}, assuming having a fixed follow-up duration in all patients; in this example the last component of the HCE is the slope of the GFR curve. If we focus on this component alone we can model the GFR measurements $Y_{ij}$,  for subject $i$ on treatment $j$  at timepoint $t$ as 

\begin{align}\label{Var2}
    Y_{ij}(t) = b_{0j}+b_{1j}t+\varepsilon_{ij}. \ \ 
\end{align}
Where \( \varepsilon_{ij} \sim \mathcal{N}(0, \sigma_e^2) \) is the measurement error, $b_{0j}$ is a random intercept, and \( b_{1j} \sim \mathcal{N}(\beta_{1j}, \sigma_s^2) \) is the random slope for  treatment $j$, $j=0,1$.

If a linear model is fitted to the data for each subject $i$, the variance of the least squares means estimate (LSME) of the slopes is given as
\begin{align}\label{eq:slopevar}
  \text{Var}(\hat{b}_{1j}) = \sigma_s^2 + \sigma_e^2\left (\frac{1}{\sum_{k=1}^n (t_k - \bar{t})^2 } \right )
\end{align} 
where \( \bar{t} \) is the mean of the time points, and \( n \) is the number of time points.
Alternatively the slope could be estimated as the mean change (MC) from a single baseline measurement to a single end of trial assessment divided by the trial duration (T). In that case equation~(\ref{eq:slopevar}) simplifies to 
\begin{align}\label{eq:slopevarcfb}
  \text{Var}(\hat{b}_{1j}) = \sigma_s^2+\sigma_e^2   \frac{2}{T^2} 
\end{align} 
The normality assumptions imply that the win probability  $\theta$ is given as  $ \Phi\left ( \frac{\hat{b}_{11} - \hat{b}_{10}  }{\sqrt{2\text{Var}(\hat{b_1})}} \right )$, where $\hat{b}_{10} - \hat{b}_{11} $  is the difference in mean slope. 

Now if we could observe the slopes for each subject without measurement error, the 'true' win probability $\theta$ could be calculated based on $\sigma^2_s$ alone. 

For these two estimation methods, the observed WR will thus be attenuated by the second term of equations~(\ref{eq:slopevar}) and~(\ref{eq:slopevarcfb}). It can be seen that the observed WR can be increased by reducing this term, that is by either decreasing the measurement error ($\sigma_e^2$), increasing the time-interval over which we measure, or for the least squares estimate by increasing the number of  observations. 

While it may initially seem appealing to fit a random slope, random intercept model, utilizing all observations, and subsequently employing individual-specific slope estimates with Best Linear Unbiased Predictions (BLUPs), it's important to note that these estimates are inherently correlated within treatment arms. This correlation violates the assumption of independent observations required by existing methods for statistical inference of WRs. This challenge will need to be resolved before BLUPs can be recommended for slope endpoints in win statistics.

In general the dependence of the WR on slope variability and study design characteristics persists regardless of whether the individual patient’s slopes are estimated by least squares (LS) or as the total GFR change from baseline to the end of follow-up measurement. 

\begin{figure}
    \centering
    \includegraphics[width=1\linewidth]{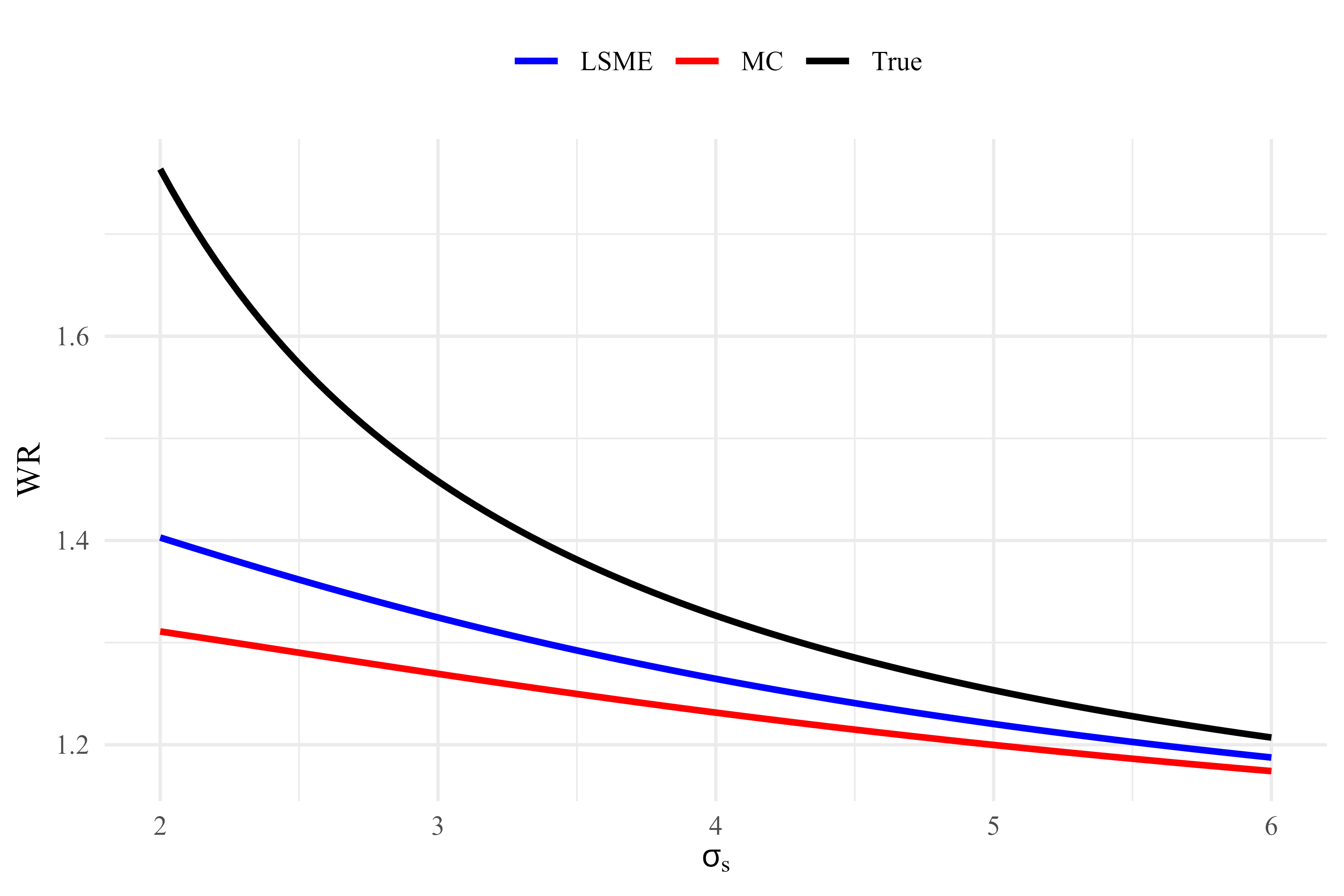}
    \caption{WR vs between-subject slope standard deviation for the LSME and MC methods, and the true value, illustrating both the bias of the estimates, and the true/estimated WR's dependence on $\sigma_s$}
    \label{fig:betweenPTSlopeSD}
\end{figure}

Figure~\ref{fig:betweenPTSlopeSD} shows how the WR depends on the population slope variability for each of the indicated analysis methods as well as for the latent true slopes for a two-year trial in which GFR is measured every 3 months, the between-patient slope SD ($\sigma_s$) is 3 ml/min/1.73m\textsuperscript{2}/yr, the treatment attenuates the mean slope from -3 ml/min/1.73m\textsuperscript{2}/yr in the control group to -2 ml/min/1.73m\textsuperscript{2}/yr in the treatment group, and the within patient error standard deviation ($\sigma_e$) is 5.18 ml/min/1.73m\textsuperscript{2}. The within patient error standard deviation is typical of past RCTs with mean baseline GFR of approximately 40 ml/min/1.73m\textsuperscript{2}, see eg \cite{Greene2019-yq}. The LSME and MC methods underestimates the WR in this case by 9\% and 13\% respectively. The WR decreases as a function of the between-patient slope standard deviation for both the true slopes and for each of the methods of slope estimation. For any given between-patient slope standard deviation, we see the expected result; both methods underestimates the compared to the 'true' slopes, and the LSME performs better than the MC method.   

\begin{figure}
    \centering
    \includegraphics[width=1\linewidth]{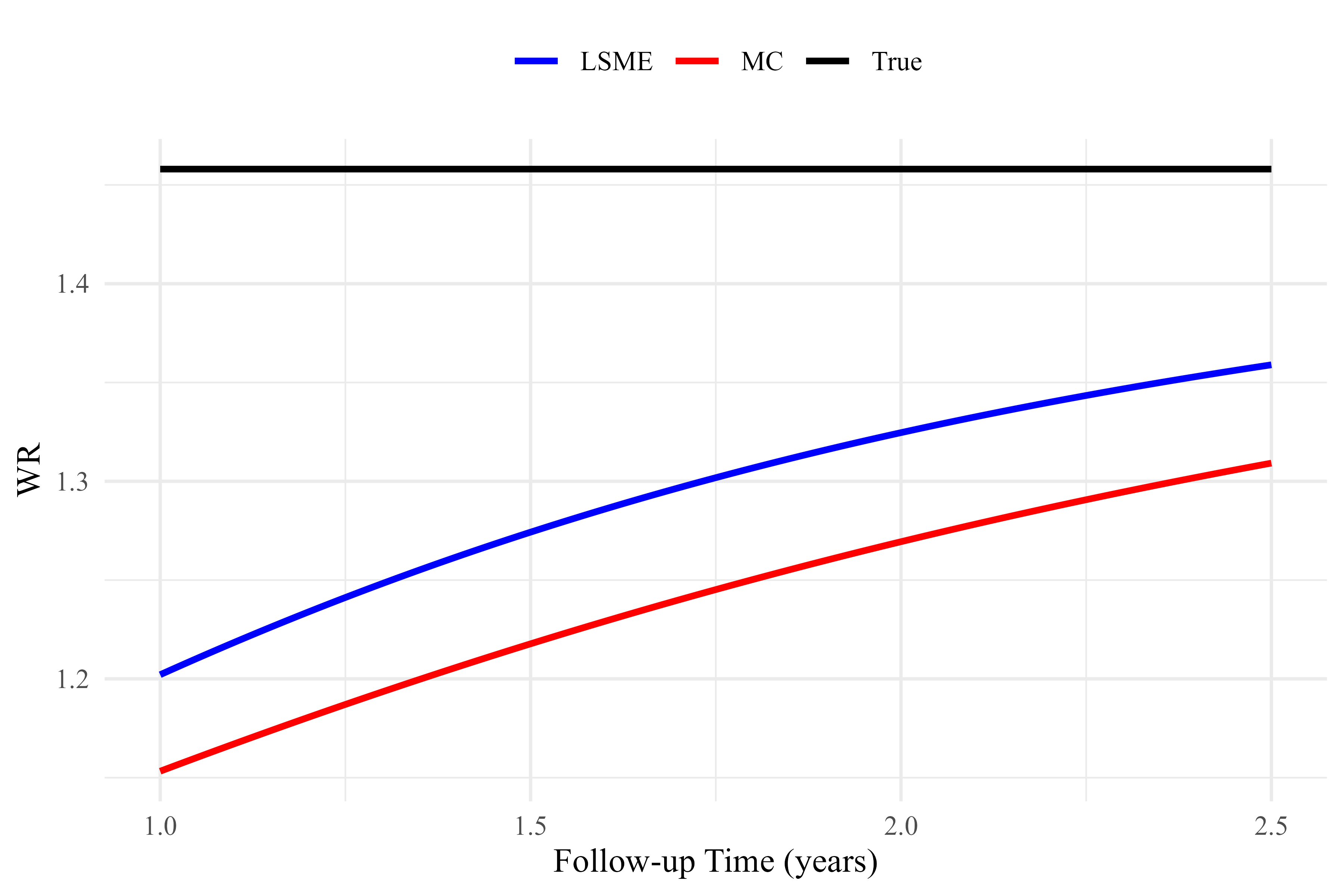}
    \caption{WR vs Follow-up time for the LSME and MC methods, and the true value, illustrating both the bias of the estimates, and the true/estimated WR's dependence on follow-up time.}
    \label{fig:WRvsFU}
\end{figure}

\begin{figure}
    \centering
    \includegraphics[width=1\linewidth]{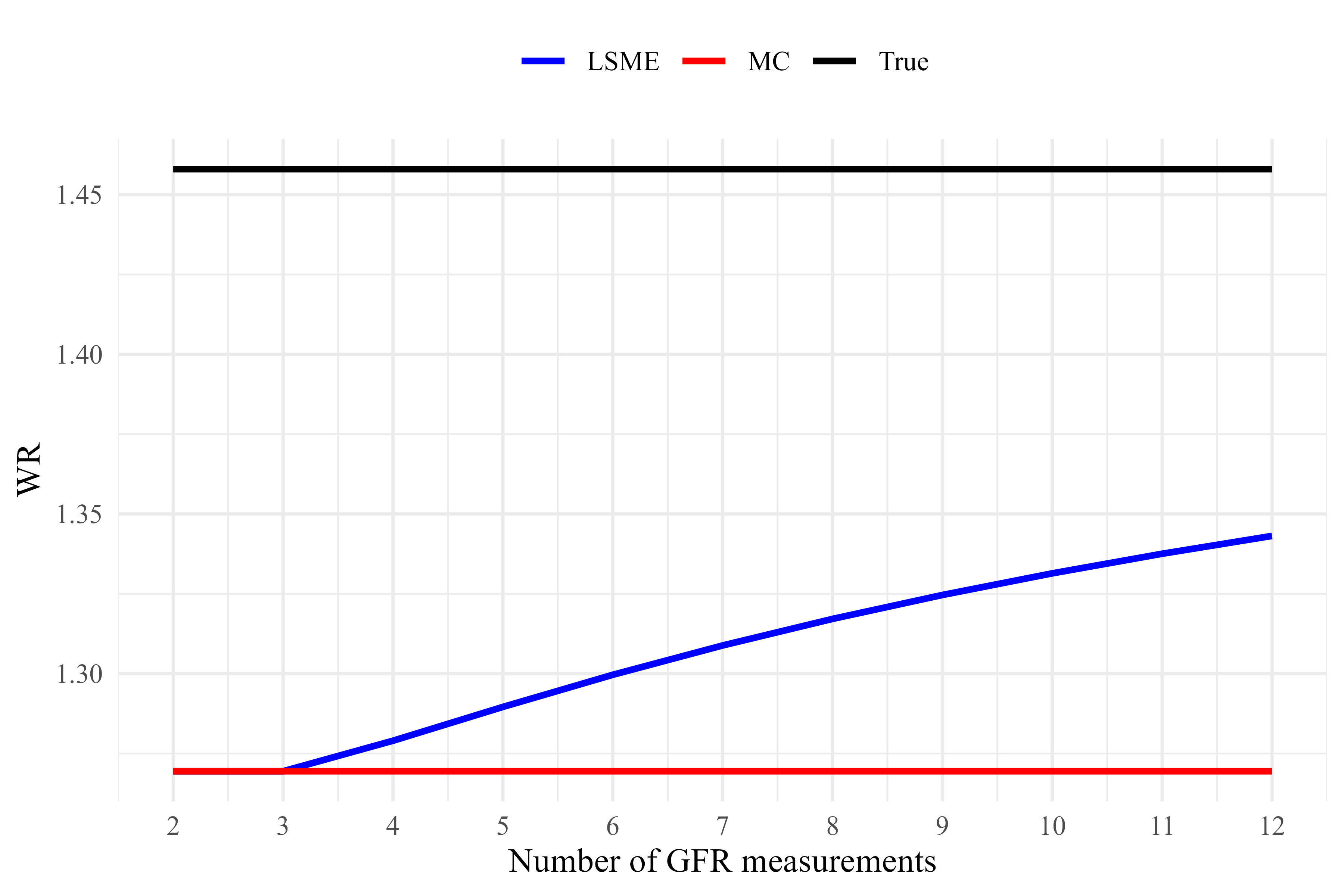}
    \caption{WR vs number of GFR measurements for the LSME and MC methods, and the true value, illustrating the bias of the estimates, and the estimated WR's dependence on the number of measurements.}
    \label{fig:WRvsNum}
\end{figure}

\noindent Figures~\ref{fig:WRvsFU} and~\ref{fig:WRvsNum} display the dependence of the WR on the duration of follow-up and on the number of measurements, respectively. The WR calculated for the latent true GFR slopes is unrelated to these design characteristics, but the WR calculated using LSME and MC increase for larger values of these design factors, and thus the bias is reduced.  

\subsection{Non-collapsibility}\label{noncollap}
Daniel et al. (2020) discussed the concept of non-collapsibility of effect measures \cite{noncollapDaniel}. In short, a treatment effect measure is said to be non-collapsible if the inclusion of a covariate in the model associated with the outcome will change the target of the treatment effect to be estimated. A non-collapsible effect measure, such as the odds ratio or the hazard ratio, will have different conditional and marginal estimates. The WR, WO, and NB are in general also non-collapsible effect measures, see eg. \cite{Buyse2025-20}. To demonstrate this concept, we provide a simple example of the win ratio that utilizes a normally distributed variable, e.g. GFR, as its sole component. 

\begin{figure}
    \centering
    \includegraphics[width=1\linewidth]{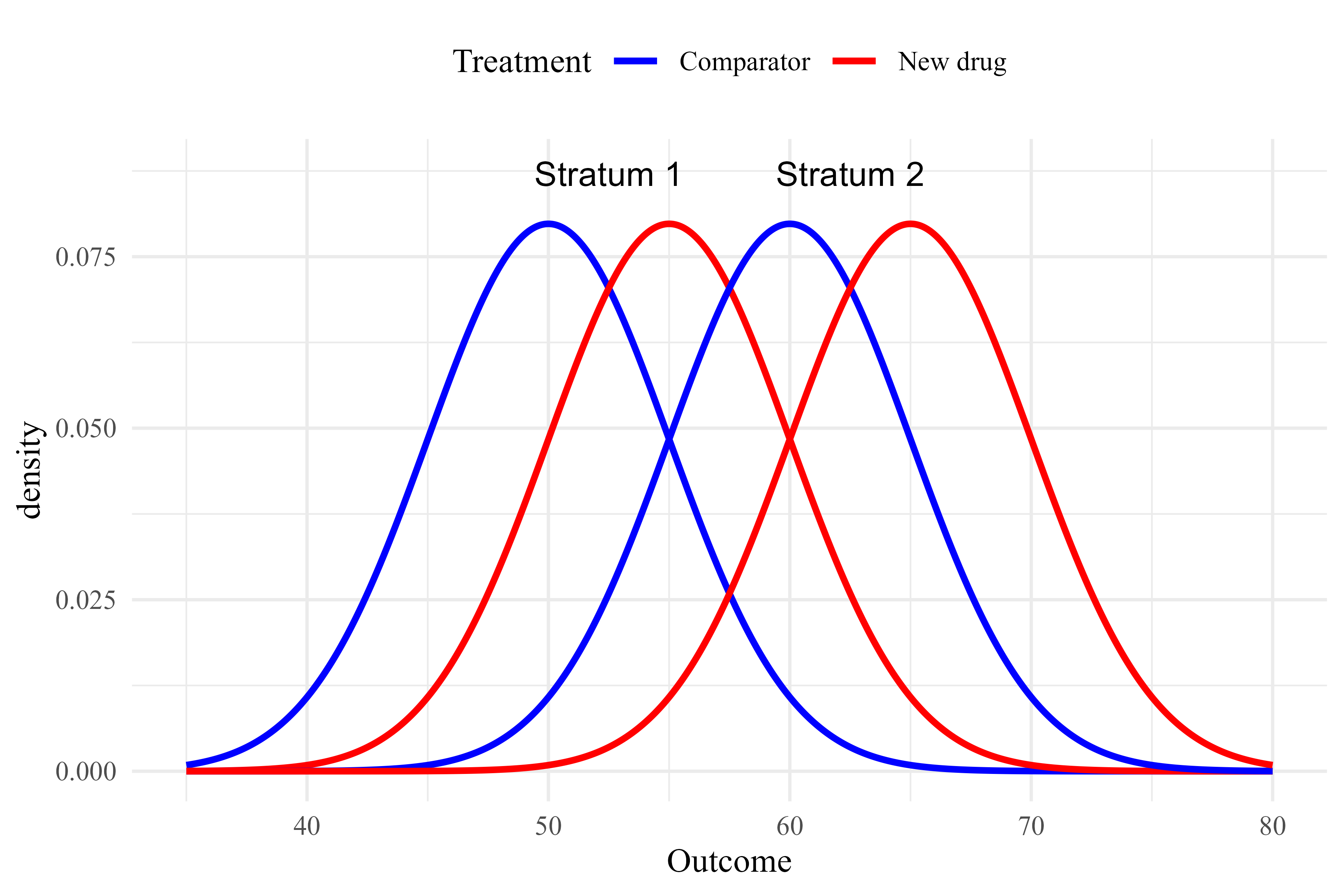}
    \caption{Normal density plots showing the distribution patterns for two treatments across two strata. Intra-strata comparisons indicate that the new drug typically shows a consistent winning trend in both strata; however, this consistency does not hold true for cross-strata comparisons}
    \label{fig:noncollap-cont}
\end{figure}

Let $Y_{sj} \sim \mathcal{N} (\mu_{sj}, \sigma_{sj}^2)$ be independent normally distributed stochastic variables, e.g. GFR changes from baseline, where $s$ denotes the stratum and $j=0,1$ denotes treatment. Further, let $Z_{sj}$ denote a strata indicator variable, such that $P(Z_{sj}=1)$ corresponds to the probability of being in stratum $s$ for treatment $j$ $ \left(\sum_s P(Z_{sj}=1)=1 \right)$. Assume that larger values are `better' and utilizing that that $Y_{sj}$'s are independent. The WR in stratum $s$ is given by 
\begin{align}  \label{eq:WRcontstrat}
    \text{WR}_s = \frac{P(Y_{s1} > Y_{s0})}{P(Y_{s1} < Y_{s0})} = \frac{P(Y_{s1} - Y_{s0} > 0 )}{1- P(Y_{s1} - Y_{s0} > 0)}
\end{align}
In contrast, the marginal WR is given by 
\begin{align*}
    \text{WR} = \frac{\sum_k \sum_l P(Z_{k1}=1, Z_{l0}=1, Y_{k1}-Y_{l0} > 0)}{1 - \sum_k \sum_l P(Z_{k1}=1, Z_{l0}=1, Y_{k1}-Y_{l0} > 0)}
\end{align*}
If we assume that strata distribution is independent of the treatment allocation, and  it is assumed that the strata are independent across the two arms, the equation is simplified to
\begin{align} \label{eq:WRcontmarg}
    \text{WR} = \frac{\sum_k \sum_l P(Z_{k1}=1) P(Z_{l0}=1) P(Y_{k1}-Y_{l0} > 0)}{1 - \sum_k \sum_l P(Z_{k1}=1) P(Z_{l0}=1) P(Y_{k1}-Y_{l0} > 0)}
\end{align}
A numerical example is illustrated in Table~\ref{tab:collap} and figure~\ref{fig:noncollap-cont}. 

The example is similar to the one given for the Odds ratio in the FDA guidance on adjusting for covariates in randomized clinical trials \cite{FDAcovar}.

In this example the population consists of two equally sized strata. The endpoint is normally distributed with population means given in the table. The standard deviation for both strata and treatments are given as $\sigma_{sj}=\sigma=5$. The strata specific WRs calculated by means of equation~(\ref{eq:WRcontstrat}) are 3.17 for both strata. The marginal WR for the entire population calculated by means of equation~(\ref{eq:WRcontmarg}) is 2.18. The distributions are illustrated in ~\ref{fig:noncollap-cont}, here it can be seen that when doing comparisons across strata the resulting WR will be "diluted" from the cross-strata comparisons.  
Note that the distribution in the `combined' population is bimodal. A stratified WR would yield the same point estimate as the strata-specific WRs, since any weighted average results in the same value. As for the odds ratio and hazard ratio, the non-collapsibility of the WR makes it problematic as a causal effect measure and complicates its transportability between different populations. This issue is particularly pronounced in meta-analyses, where combining results from different studies becomes challenging. More details on causality and non-collapsibility can be found in \cite{Didelez2021-nc}.

\begin{table}[ht]
    \centering
    \begin{tabular}{lcccc}
        \toprule
                        & \textbf{Percentage of} & \multicolumn{2}{c}{\textbf{Mean}}& \textbf{WR} \\ 
                        & \textbf{population}    &          \textbf{New drug} & \textbf{Comparator} &    \\ 
        \midrule
        Strata 1 & 50\%       & 55& 50& 3.17\\ 
        Strata 2 & 50\%       & 65& 60& 3.17\\ 
        Combined           & 100\%      &          -&            -& 2.18\\ 
        \bottomrule
    \end{tabular}
    \caption{Non-collapsibility of the WR in a hypothetical target population for a normally distributed response in each stratum, with means given in the table and a common standard deviation of 5. Means for the combined case are not displayed due to the bimodal distributions.}
    \label{tab:collap}
\end{table}

\section{Discussion}\label{disc}

In this study, we have explored some of the theoretical and practical implications of using the Win Ratio (WR) in the context of Hierarchical Composite Endpoints (HCEs) in clinical trials, particularly with examples from chronic kidney disease (CKD). While the WR offers a structured way to compare treatments, several challenges and limitations complicate its interpretability, particularly regarding its relationship with the estimand framework, causal interpretation, transitivity, variance dependence, and its non-collapsible nature. 

The application of win statistics for HCEs as primary analyses has gained acceptance as indicated by two cases of FDA approvals for cardiac amyloidosis (Tafamidis \cite{maurer2018tafamidis},  ATTR-ACT ClinicalTrials.gov number, \href{https://clinicaltrials.gov/study/NCT01994889}{NCT01994889}. and Acoramidis \cite{Gillmore2024-iz},  ATTRibute-CM ClinicalTrials.gov number, \href{https://clinicaltrials.gov/study/NCT03860935}{NCT03860935} ). However, it is worth noting that in the case of Tafamidis, there is no mention of the win statistics, only the p-value of the underlying Finkelstein-Schoenfeld method.

The use of WR in HCEs raises several theoretical concerns. One foundational issue arises from defining an appropriate estimand for WR-based analyses. According to the ICH E9 addendum on estimands, the estimand must address the clinical question being investigated while ensuring clarity and interpretability. 
The WR, while useful in certain contexts, does not always provide a clear interpretable measure of the treatment effect. However, in some trials, it may be infeasible to address the most relevant clinical questions, such as overall long-term survival, necessitating compromises that hinder rigorous adherence to the principles of the estimand. 

The WR is a population-level effect measure, which compares the outcomes of different subjects across treatment groups. This can lead to different conclusions than the more natural, but non-identifiable, individual-level effect measures, which compare potential outcomes within the same subject. Hand's paradox illustrates this discrepancy between these two. The toy example presented in table~\ref{tab:handsparadox} is intentionally crafted to highlight the distinction between these. It is extreme in that one patient experiences a significant detrimental effect from the active treatment, while the two other patients experience only a minor beneficial effect. In this scenario one could question the relevance of any overall effect estimate. The main point of this example is to demonstrate that the individual-level measure provides a more relevant causal measure; however, this is not identifiable, and the two measures are not guaranteed to be even similar. Recently, \cite{Even2025-hr} suggested that a matching approach based on a nearest neighbor algorithm, which can be seen as an extreme form of stratification, could be a way to approximate the non-identifiable individual-level effect measure. 

Non-transitivity is an additional concern, particularly in multi-arm trials or when comparing treatments indirectly. The WR may indicate that treatment A is better than B, B better than C, and yet C better than A, making it impossible to establish a consistent treatment ranking. This paradox highlights the need for caution when interpreting WR results in comparative effectiveness research. 

Another important limitation is the WR’s dependence on the variance of continuous components. As shown in our slope example, the magnitude of the WR can be heavily influenced by measurement error and study design characteristics such as follow-up duration and frequency of measurements. This sensitivity reduces the WR’s utility as a pure effect measure and instead positions it more as a measure of discrimination between treatment groups. 

An additional key concern is the WR’s non-collapsibility, a property shared with other widely used metrics such as odds ratios and hazard ratios. The non-collapsibility of the WR complicates causal interpretation and can lead to inconsistencies when comparing results across trials or conducting meta-analyses. This issue becomes particularly critical when the population differs significantly. 

Although the WR holds promise in its ability to evaluate complex composite endpoints, the aforementioned limitations underscore the importance of complementary analyses. Key recommendations include: (1) defining fixed follow-up periods; (2) dissecting individual HCE components and employing visualizations to clarify the drivers of observed WR values, see e.g. \cite{Karpefors2023-sc}; and (3) avoiding the combination of components with vastly differing levels of severity. 

In summary, while the WR introduces an innovative and versatile framework for summarizing multifaceted outcomes, its limitations must be carefully considered. We propose considering WR as a discriminatory measure between distributions in active and control groups, similar to non-parametric tests. One such application is being presented in \cite{Hartley2025-zj} who use win statistics to improve commit-to-Phase-3
decision-making in oncology. 

\section{Acknowledgment}
The authors would like to acknowledge the use of generative AI (ChatGPT, OpenAI o3 Mini) for language enhancements in the preparation of this work. 

\bibliographystyle{unsrt} 
\bibliography{bibliography}

\end{document}